\documentclass[preprint,aps,12pt,showpacs,nofootinbib,tightenlines]{revtex4}
\usepackage{amsmath}
\usepackage{amssymb}
\usepackage{epsfig}
\usepackage{graphicx}
\usepackage{ctex}
\begin{document}
\def\pslash{\rlap{\hspace{0.02cm}/}{p}}
\def\eslash{\rlap{\hspace{0.02cm}/}{e}}
\def\LL{\cal L}

\title {Associated production of the heavy charged gauge boson ${W_{H}}$ and a top quark at LHC }
\author{Qing-Guo Zeng$^{1,2}$}
\author{Shuo Yang$^{3}$$^{\dag}$}
\author{Chong-Xing Yue$^{1}$$^{*}$}
\author{You Yu $^{1}$}
\address{%
\small
 $^1$ Department of Physics, Liaoning Normal University, Dalian 116029,  China\\
$^2$ Department of  Physics, Shangqiu Normal University, Shangqiu  476000,  China\\
$^3$ Physics Department, Dalian University, Dalian, 116622,  China
    \vspace*{1.5cm}}
\thanks{%
  cxyue@lnnu.edu.cn \\
$ ^{\dag}$ yangshuo@dlu.edu.cn}
\begin{abstract}
In the context of topflavor seesaw model, we study the production of the heavy charged
 gauge boson ${W_{H}}$ associated with a top quark at the LHC.
Focusing on the searching channel $pp\rightarrow tW_H\rightarrow t\bar{t}b \rightarrow l\nu jjbbb$, we carry out a full simulation
of the signal and the relevant standard model  backgrounds.
The kinematical distributions of final states are presented.
It is found that the backgrounds can be significantly suppressed by sets of kinematic cuts, and the signal of the heavy charged boson might be detected
at the LHC with $\sqrt{s}=14$ TeV. With a integrated luminosity of $\LL= $ 100 $fb^{-1}$, a $8.3 \sigma$ signal significance can be achieved for $m_{W_H}=1.6$ TeV.

\end{abstract}
 \maketitle
\section{ Introduction}
\noindent

The standard model (SM) of particle physics is one of the most successful theories
over the past decades which describes a variety of experimental results.
However, the theoretical shortcomings of the SM, such as quadratic
divergencies, the triviality of a $\phi^4$ theory, etc., suggest that it should be embedded in a larger
scheme. Many popular new physics models beyond the SM have been proposed, and some
of which predict the existence of the new gauge bosons
with masses at the TeV order. Thus, search for extra gauge bosons provides a common tool in quest for new
physics at the LHC. In this paper, we study the signal of the heavy charged gauge boson $W_H$ in topflavor seesaw model at the LHC.

It is interesting to note that only the top quark mass lies at the same mass scale of
weak gauge bosons, while all other fermions are provided with masses less than a few GeV. This suggests that the top
quark sector may involve some new gauge dynamics at the weak scale in contrast to all light fermions. Topcolor seesaw models employ strong $SU(3)$ top gauge group with singlet heavy quarks to address this mass hierarchy in fermion sector~\cite{topcolorseesaw}. As an alternative, the topflavor seesaw model was proposed~\cite{9911266}.
In topflavor seesaw model, the top sector experiences a new ${SU(2)_{t}}$ gauge interaction (Type I) \footnote{The topflavor seesaw models could be implemented by introducing additional $SU(2)_{t}$ (TypeI) or $U(1)_{t}$ (Type II) gauge group\cite{9911266}. In this paper, we only focus on the Type I realization.}.
The gauge group of topflavor seesaw model is ${SU(3)_{c}}\otimes{SU(2)_{t}}\otimes{SU(2)_{f}}\otimes{U(1)_{y}}$.
In this model, the first two family fermions are singlets under the new ${SU(2)_{t}}$ gauge interaction.
At the same time, a doublet of spectator quarks $S=(T,B)^{T}$ is introduced to cancel the theory anomaly for the third family.
Two Higgs doublets $\Phi_{1}$ and $\Phi_{2}$ are introduced to spontaneously break gauge group ${SU(3)_{c}}\otimes{SU(2)_{t}}\otimes{SU(2)_{f}}\otimes{U(1)_{y}}$ down to the gauge subgroup  $SU(3)_{c}\otimes U(1)_{em}$.
Thus, two neutral physical Higgs boson $h^{0}$
and $H^{0}$ are predicted, in addition to the heavy bosons $(W_{H}, Z_{H})$.
It is exciting that LHC has discovered  a Higgs-like boson with mass around 125 GeV~\cite{ATLAS,CMS}, which coincides with the light Higgs  $h^{0}$ predicted in the topflavor seesaw model~\cite{1304.2257}.

In topflavor seesaw model, in order to address the mass hierarchy in fermions, only the top-sector enjoys the extra $SU(2)_t$ gauge interaction which is stronger than the ordinary $SU(2)_f$ (associated with all the other light fermions). After the gauge symmetry breaking $SU(2)_t\times SU(2)_f \rightarrow SU(2)_L$ at a high scale $u$, the heavy gauge bosons which are combinations of the corresponding broken gauge fields of $SU(2)_t$ and $SU(2)_f$ obtain mass. This results in the coupling of $W_H$ with $tb$ is enhanced by the mixing parameter $1/x$ while the couplings of $W_H$ with light fermions are suppressed by a factor of $x$~\cite{1304.2257}. Here, $x$ is the ratio of the gauge coupling $g_1$ of $SU(2)_f$ to the gauge coupling $g_0$ of $SU(2)_t$ and it is constrained to be $x^{2}\ll 1$ in this construction.
This feature of topflavor seesaw model is different from those in many new physics model with heavy $W_H$ including Kaluza-Klein theories of extra dimensions - as the excitation of the $W$\cite{exd}, little Higgs theories - as the gauge bosons of the extended symmetry\cite{lh}, several other well-motivated extensions of the SM\cite{cqh,1111.5021,g1312,1011.5918,1111.1551}.
In this paper, we study the production of $W_H$ associated with a top quark
followed by the decay $W_H \rightarrow tb$. This process provides an independently test of the $W_H\bar{t}b$ coupling which is closely relevant to the model feature.

This paper is organized as follows. In section II, we give a brief
review of topflavor seesaw model and show relevant couplings for our calculation. In section III, the phenomenological analysis and numerical calculations for the production of the heavy charged gauge boson  $W_H$ in association with a top quark are presented. Our conclusions are given in section IV.

\section{ A brief review of topflavor seesaw model}
 \noindent

The topflavor seesaw model~\cite{9911266}, in which the top sector experiences a new ${SU(2)_{t}}$
gauge interaction, is based on gauge group
${SU(3)_{c}}\otimes{SU(2)_{t}}\otimes{SU(2)_{f}}\otimes{U(1)_{y}}$.
 The corresponding three gauge couplings of the gauge group
${SU(2)_{t}}\otimes{SU(2)_{f}}\otimes{U(1)_{y}}$ are
${g_{0}},{g_{1}},{g_{2}}$.
 Two Higgs doublets $\Phi_{1}$ and $\Phi_{2}$ are invoked to spontaneously break
${SU(3)_{c}}\otimes{SU(2)_{t}}\otimes{SU(2)_{f}}\otimes{U(1)_{y}}$
down to the residual symmetry ${SU(3)_{c}}\otimes{U(1)_{em}}$.
The Higgs doublet $\Phi_{1}$ with a nonzero vacuum expectation
value (VEV) ${u }$ break ${SU(2)_{t}}\otimes{SU(2)_{f}}$ down
to the SM gauge group ${SU(2)_{L}}$, and the Higgs doublet
$\Phi_{2}$ with a VEV ${\upsilon }$ break
${SU(2)_{L}}\otimes{U(1)_{y}}$ down to the SM gauge group
${U(1)_{em}}$. This kind of breaking pattern makes that this model contains extra massive color-singlet heavy
gauge bosons ${W_{H}}$ and ${Z_{H}}$, in addition to two neutral
physical Higgs boson ${h^{{0}}}$ and ${H^{{0}}}$.

The structure of topfavor seesaw model including its Higgs, gauge and top sectors has been systematically studied
in Refs. \cite{9911266,1304.2257}.
 In this work, we aim to study the production of the
 heavy charged boson ${W_{H}}$ at the LHC. The couplings of the  heavy charged boson ${W_{H}}$
 to ordinary particles, which are related to our calculation, are given
 by \cite{1304.2257},

\begin {equation}
\label{eqLargff} {\cal L_{W_{H}F\bar{F'}}} =
\frac{-ixg}{\sqrt{2}}\overline{f}\gamma_{\mu}P_L V_{\bar{f}f'}f'W^{+}_{H}
+\frac{i(1-rx^{2})g}{\sqrt{2}x(1+r)}\overline{t}\gamma_{\mu}P_LbW^{+}_{H}
+h.c,
\end {equation}
and
\begin {equation}
\label{eqLargH} {\cal L_{H}}=
-ix(c_{\alpha}-ys_{\alpha})gm_{_{W}}W^{+}_{H}W^{-}h^{0}+ix(s_{\alpha}+yc_{\alpha})gm_{_{W}}W^{+}_{H}W^{-}H^{0},
\end {equation}
where we have defined   the ratios $x \equiv g_{1}/g_{0} $, $r\equiv k^{2}/M^{2}_{S} $, and $y \equiv \upsilon/u$.
Here $ M_{S} $ denotes the  mass eigenvalues parameter of heavy quarks $(T,B)$, and the parameter $ k$ is expected to be of $\mathcal{O}(M_{S})$.
And $\alpha $ represents the mixing angle between physical Higgs bosons ($h^{0}, H^{0}$) and their weak eigenstates.
The Higgs-like particle found by LHC agrees well with the SM prediction which suggests a small $\alpha$ in the range $0< \alpha \leq 0.2 \pi $ \cite{ATLAS,CMS,1304.2257}.

The coupling of ${W_{H}}$ with light gauge bosons $W$ and $Z$ is
given by
\begin {equation}
\label{eq3} {g_{W_{H}WZ}}=
\frac{-ix^{3}y^{2}}{c^{2}_{W}}G^{SM}_{WWZ},
\end {equation}
where $G^{SM}_{WWZ}$ represents the coupling
constant of $WWZ$ in the SM.

A detailed analysis of direct search for the topflavor seesaw model at the LHC and the  constraints on this model of electroweak precision measurements are presented in Ref.~\cite{1304.2257}.
In the topflavor seesaw model, the heavy spectator quarks $(T,B)$ are vector-like under $SU(2)_L$, so the fermionic contributions to oblique corrections can be fairly small in decoupling limit.
The higgs sector contributions are relevant to the masses of the physical Higgs bosons $(h^0, H^0)$ and the mixing angle $\alpha$.
The contributions from gauge sectors are dependent on the masses of heavy gauge bosons $W_H/Z_H$ and the mixing parameter $x$.
After deducing the contributions to electroweak precision parameters from gauge, Higgs and fermion sectors,
it is found that at $95\%$ C.L., there should be  $m_{W_{H}}\gtrsim 0.45 \sim 1$ TeV for a wide $H^0$ mass range up to 800 GeV
with the inputs $\alpha = 0.1\pi \sim 0.2\pi$ and $M_T = 4$ TeV \cite{1304.2257}.

The ATLAS and CMS collaborations have been actively searching for the new gauge boson $W_{H}$
at the LHC \cite{CMSW',ATLASW'}. They mainly focus on the sequential standard model (SSM),
where the couplings of $W_{H}$ with fermions equal to the corresponding SM couplings.
Focusing on the process $pp \rightarrow W_{H} \rightarrow l\nu$, the lower limits for the $W_{H}$ mass of 2.15 TeV \cite{ATLASW'lv}
and 2.5 TeV \cite{CMSW'lv} have been obtained at the LHC by the ATLAS and
CMS experiments respectively. CMS has also searched for the process $pp\rightarrow W_{H} \rightarrow WZ$ using the
fully leptonic final state and has set the lower limit for $W_{H}$ bosons $m_{W_{H}} > 1.14 $ TeV \cite{CMSW'WZ}.
However, different scenarios have different phenomenological features. The direct search constraints can be relaxed in some models.
In the topflavor seesaw model, the  couplings of heavy bosons $W_H/Z_H$ with light fermions are suppressed by the small mixing angles, which are of the order of $\mathcal{O}(x)$. Hence, the production rate for the process $pp \rightarrow W_H$ is suppressed by a factor of $x^2$ and the decay rates for $W_H\rightarrow f\bar{f'}$ and
$W_{H} \rightarrow  WZ$ are also suppressed. Thus, the corresponding signals are hard to be detected.

The most stringent limit on the $W_H$ mass comes
from $pp\rightarrow W_{H} \rightarrow tb$ given by the CMS collaboration \cite{CMSW'}.  CMS has excluded right handed  $W_{H}$ with mass below $1.85$ TeV via this channel. Unlike SSM model, the couplings of $W_H$ to light fermions are suppressed by $x$ while the coupling of $W_Htb$ is enhanced by a factor $1/x$ in the topflavor seesaw model. Considering the branching ratios, the $W_H$ signal rate of $\sigma \times Br$ is smaller than that of SSM by about a factor of $4x^2$.
With the sample input $x = 0.15$, it is found a 95\% C.L. lower limit on $W_H$ mass, $m_{W_H} > 1.0 $ TeV, from the CMS data \cite{1304.2257}\cite{CMSW'lv,CMSW'WZ,CMSW'}.

\section{Numerical results and discussion}
\begin{figure}[htbp]
\includegraphics [scale=0.80] {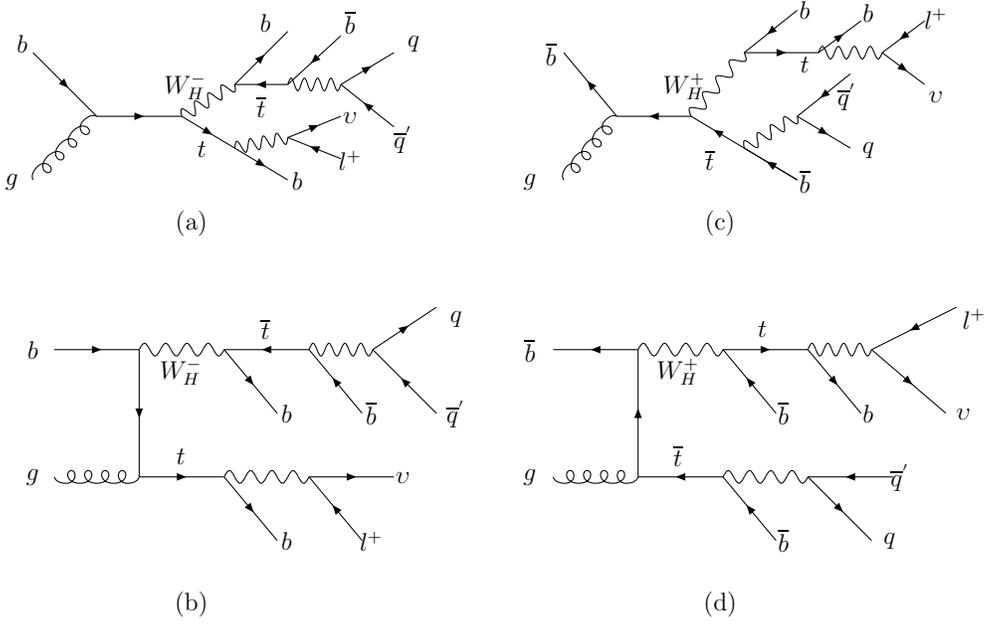}
\caption{ The partonic level process for a $W_{H}$  in association with a top quark  production and decay in
hadronic collisions.}
  \label{Eqs:fm}
\end{figure}

\noindent

Before studying the process  $pp\rightarrow tW_{H}$, we
firstly consider the possible decay modes of the heavy charged gauge boson
$W_{H}$.
Once produced at the LHC, the heavy  boson ${W_{H}}$ can decay into $f\bar{f}'$, $t\bar{b}$, $Wh^{0}$, $WH^{0}$ and $WZ$.
The partial widths of $W_{H}$ decaying to a pair of fermions are
\begin {equation}
\label{eq4}\Gamma(W_{H}\rightarrow f\bar{f^{'}})=\frac{g^{2}x^{2}}{16\pi}\mid V_{f\bar{f'}}\mid^{2}m_{W_{H}},
\end {equation}
\begin {equation}
\label{eq5}\Gamma(W_{H}\rightarrow
t\bar{b})=\frac{g^{2}}{16\pi}[\frac{1-rx^{2}}{x(1+r)}]^{2}m_{W_{H}}
(1-\frac{m_{t}^{2}}{m_{W_{H}}^{2}})(1-\frac{m_{t}^{2}}{2m_{W_{H}}^{2}}-\frac{m_{t}^{4}}{2m_{W_{H}}^{4}}).
\end {equation}
  Likewise, the partial widths of
the $W_{H}$  decaying to W boson and Higgs are,
\begin {eqnarray}
\label{eq6}\Gamma(W_{H}\rightarrow
Wh^{0})=\frac{g^{2}[x(c_{\alpha}-ys_{\alpha})]^{2}}{192 \pi m^{5}_{W_{H}}m^{2}_{W}}
((m^{2}_{W_{H}}+m^{2}_{W}-m^{2}_{h^{0}})^{2}-24m^{2}_{W_{H}}m^{2}_{W})\nonumber\\
\times\sqrt{(m^{2}_{W_{H}}-(m^{2}_{W}+m^{2}_{h^{0}})^{2})(m^{2}_{W_{H}}-(m^{2}_{W}-m^{2}_{h^{0}})^{2})},
\end {eqnarray}
\begin {eqnarray}
\label{eq7}\Gamma(W_{H}\rightarrow
WH^{0})=\frac{g^{2}[x(s_{\alpha}+yc_{\alpha})]^{2}}{192 \pi m^{5}_{W_{H}}m^{2}_{W}}
((m^{2}_{W_{H}}+m^{2}_{W}-m^{2}_{H^{0}})^{2}-24m^{2}_{W_{H}}m^{2}_{W})\nonumber\\
\times\sqrt{(m^{2}_{W_{H}}-(m^{2}_{W}+m^{2}_{H^{0}})^{2})(m^{2}_{W_{H}}-(m^{2}_{W}-m^{2}_{H^{0}})^{2})}.
\end {eqnarray}
 Noting the coupling ${g_{W_{H}WZ}}\propto(x^{3}y^{2})$,  and $x^{2}, y^{2}\ll1$,
  we neglect the $W_{H}\rightarrow WZ$ decay channel here.


Here, we calculate the branching ratios (BRs) of $W_H$. It is found that the BR of $W_H\rightarrow tb$ is dominant and the BRs of other decay modes $W_H\rightarrow ff'$, $W_H\rightarrow Wh^0$ and $W_H\rightarrow WH^0$ are tiny.
This is because the coupling of $W_Htb$ is roughly proportional to $1/x$ while the other decay modes are suppressed by the relevant couplings.
\begin{figure}[htbp]
\begin{center}
\includegraphics [scale=0.85] {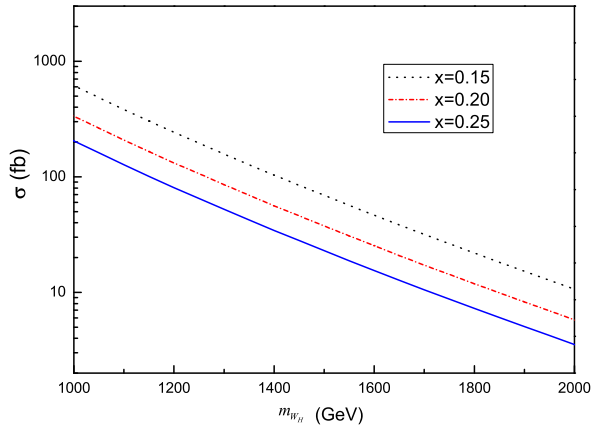}
\vspace{-0.2cm}\caption{The cross section for the process $pp \rightarrow
tW_{H}$ as function of the heavy boson mass ${m_{W_{H}}}$
for $r=1$ and three typical values $x$ at the LHC with $\sqrt{s}=14 $ TeV. }\label{Eqs:1jiemian}
\end{center}
\end{figure}

Furthermore, we show the cross section for process $pp \rightarrow tW_H$ ( include both
$g\bar{b} \rightarrow \bar{t}W^{+}_{H}$  and $ gb \rightarrow tW^{-}_{H}$ ) as a function of the heavy boson mass ${m_{W_{H}}}$ with $\sqrt{s}=14$ TeV in Fig.~\ref{Eqs:1jiemian}.
In calculations, we use the parton
distribution functions given by  CTEQ6L1 \cite{cteq}. The renormalization scale $\mu_{R}$ and the factorization scale
$\mu_{F}$ are chosen to be $\mu_{R}=\mu_{F}=m_{W_{H}}$, and the
strong coupling constant $\alpha_{S}$ is taken as
$\alpha_{S}(m_{Z})=0.118$. The SM input parameters are taken as  $m_{t} = 173.2 GeV $ and $ S^{2}_{w}=0.231 $ \cite{sm}.
As shown in the Fig. 3,  the cross sections
 decrease as $m_{W_{H}}$ increases, which are in the ranges of hundreds $fb$ to several $fb$, for $m_{W_{H}}= 1000GeV \thicksim 2000GeV$ and $x=0.15\thicksim0.25$.
For a typical mass $m_{W_H}=1.6$ TeV ($x=0.2$), there will be about 2500 $tW_H$ events at the LHC with a integrated luminosity 100 $fb^{-1}$.
The couplings of $W_{H}$  with $t\bar{b}$
is enhanced by about a factor of $1/x$ in topflavor seesaw model.
Hence, the cross section for process $pp \rightarrow tW_{H}$ in topflavor seesaw model is
larger than that in SSM, top-philic $W'$ model \cite{cqh}, little Higgs models \cite{ylh}, and left-right twin Higgs model \cite{lrth}.

In this paper, we will focus on the $tW_H$ production followed by the dominant decay channel $W_{H}\rightarrow t\bar{b}$.
And we demand a hadronic decay of
the antitop $\bar{t}\rightarrow \bar{b}W^{-} \rightarrow \bar{b}jj$ and leptonic decay of the
top quark $t \rightarrow bW^{+}\rightarrow bl^{+}\nu_{l} $ where the charged lepton provides detector trigger \footnote{Ref.~\cite{shuo} finds that the signal for heavy charged gauge boson could be extracted in full hadronic mode with the help of the jet substructure technique.}.
Thus, in the following, we investigate the signal processes

\begin {equation}
\label{eq8}pp \rightarrow g\bar{b} \rightarrow \bar{t}W^{+}_{H}\rightarrow
\bar{t}t\bar{b}\rightarrow W^{-}\bar{b}t\bar{b} \rightarrow
jj\bar{b}l^{+}\nu b\bar{b},
\end {equation}
\begin {equation}
\label{eq9}pp \rightarrow gb \rightarrow tW^{-}_{H}\rightarrow t\bar{t}b\rightarrow
W^{+}b\bar{t}b \rightarrow l^{+}\nu bjjb\bar{b}.
\end {equation}
Both electrons and muons are considered for the positive charged
lepton in our analysis. Thus the signal includes a
isolated charged lepton, five jets, and a large missing transverse momentum
$\rlap/E_{T}$ from the missing neutrino. The Feynman diagrams for the signal processes are shown in Fig.~\ref{Eqs:fm}.
MadGraph/MadEvent \cite{mad} is adapted to generate both the signal and background processes.

For the $bbbjjl\nu $ signal, the SM backgrounds mainly
  come from the irreducible $t\bar{t}b$ background and the reducible $t\bar{t}j$ background,
   where the light jet $j$ means the light-flavor quarks or gluons.
\begin {equation}
\label{eq10}pp\rightarrow t\bar{t}b \rightarrow b\bar{b}bjjl^{+}\nu,
\end {equation}
\begin {equation}
\label{eq11}pp \rightarrow t\bar{t}j\rightarrow b\bar{b}jjjl^{+}\nu.
\end {equation}
The other SM background processes, eg, $Wjjjjj$, $WWjjj$ and
$WZjjj$, etc. can be dramatically reduced by the cuts
adopted in the following and therefore we neglected them here.

In order to identify the isolated jet (lepton), we define the
angular separation between particles $i$ and  $j$ as
\begin{eqnarray}
  \label{eq12}\Delta R_{ij}\equiv\sqrt{(\Delta\phi_{ij})^{2}+(\Delta\eta_{ij})^{2}},
\end{eqnarray}
where $\Delta\phi_{ij}=(\phi_{i}-\phi_{j})^{2}$ and
$\Delta\eta_{ij}=(\eta_{i}-\eta_{j})^{2}$. $\eta_{i}(\phi_{i})$
denotes the rapidity (azimuthal angle) of the related lepton (jet).

 The basic acceptance cuts, referred to as cut I, are applied for the signal and background
events,
\begin{eqnarray}
 p_{Tj}\geq 25GeV,~~~~~~~~~~~~~~~~|\eta_{j}|\leq2.5,\nonumber\\
 p_{Tl}\geq 25GeV,~~~~~~~~~~~~~~~~|\eta_{l}|\leq2.5,\nonumber\\
~~~~\Delta R_{jj,jl}\geq0.4,~~~~~~~~~~~~\rlap/E_{T}\geq25GeV .
\label{eq13}
\end{eqnarray}
Here, $p_{Tj}$ $(p_{Tl})$   is  the jet (lepton) transverse
momentum, $\rlap/E_{T}$ denotes the missing transverse momentum from
the invisible neutrino in the final state.
The effects of these cuts are shown in Tables I and II.

To make our analysis more realistic, we simulate detector resolution effects by smearing the
lepton and jet energies according to the assumed Gaussian resolution parametrization
\begin {equation}
\frac{\delta(E)}{E}=\frac{a}{E}\oplus b,
\label{eq14}
\end {equation}
where $\frac{\delta(E)}{E}$ is the energy resolution, $a$ is a sampling term, $b$ is a constant term, and $\oplus$ denotes
a sum in quadrature. We take $a = 5\%$ and $ b = 0.55\%$ for leptons, and take $a = 100\%$ and $ b = 5\%$ for
jets \cite{smear,tag}.

\begin{figure}[htbp]
\begin{center}
\includegraphics [scale=0.396] {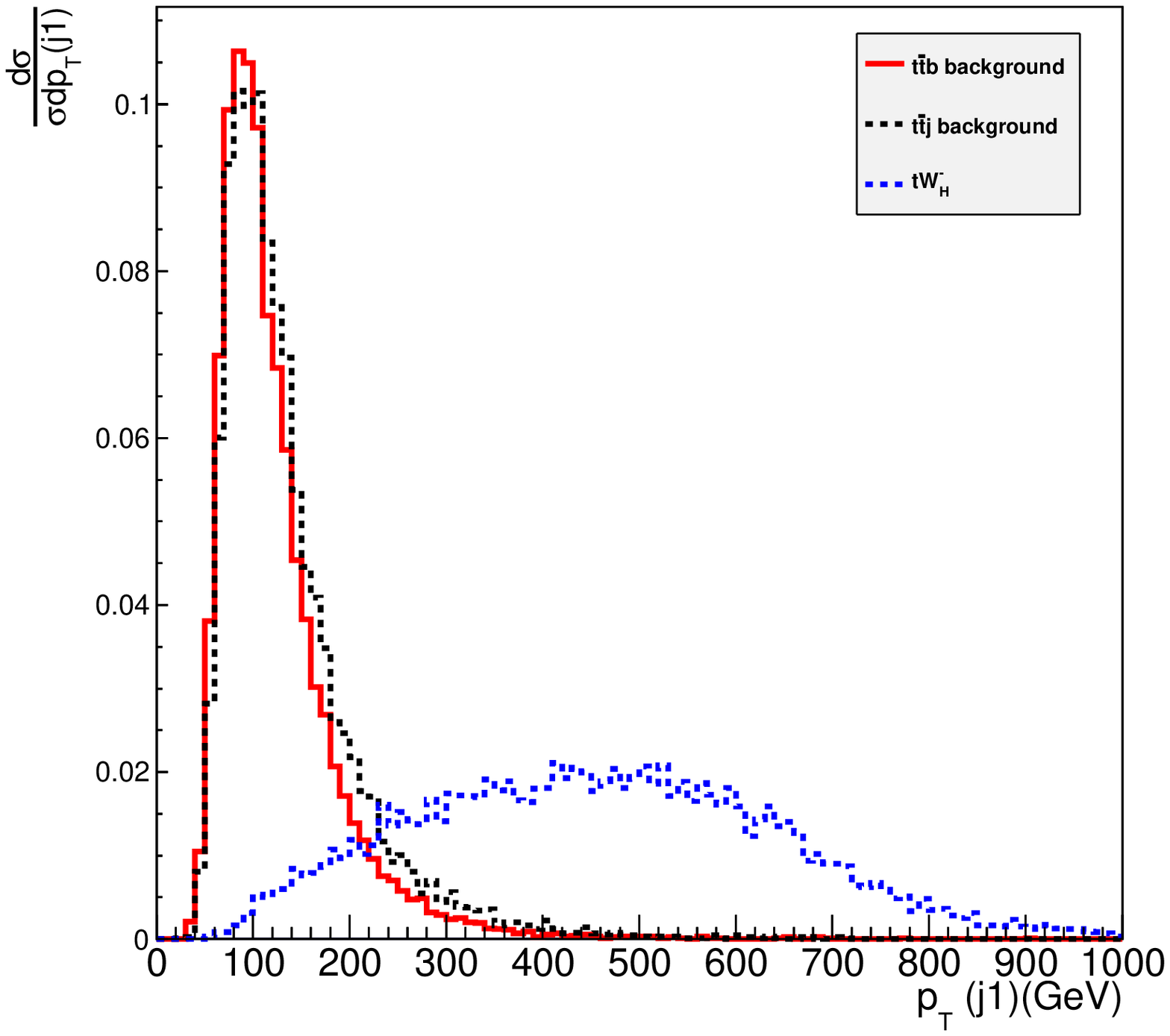}
\includegraphics [scale=0.396] {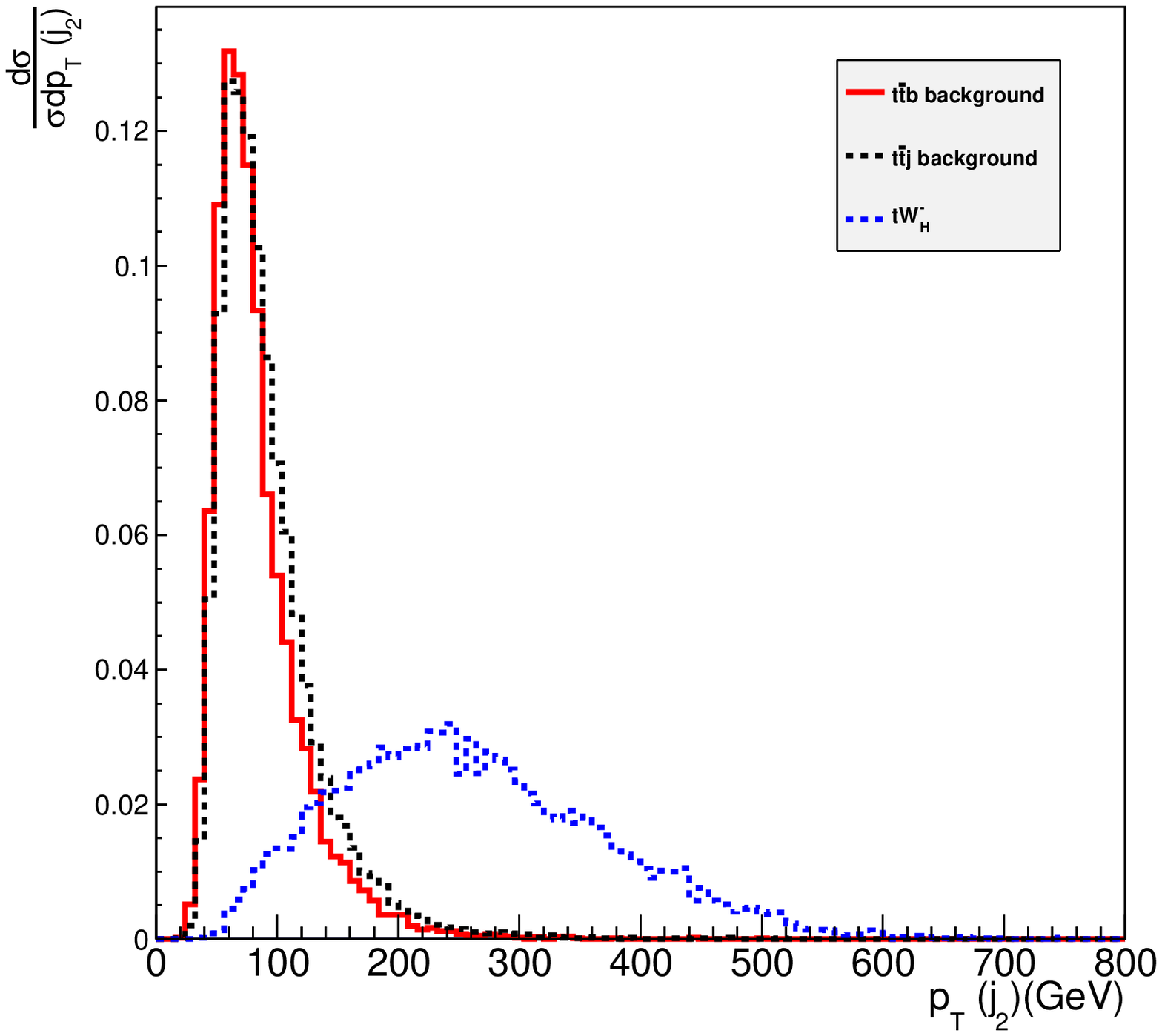}
\vspace{-0.2cm}(a)~~~~~~~~~~~~~~~~~~~~~~~~~~~~~~~~~~~~~~~~~~~~~~~~~~~~~~~~~~~~~~~~~~~~~~~~~~~~~~~~~~(b)
\vspace{-0.2cm}\caption{
Transverse momentum $p_{T}$ distributions of the hardest jet (a) and the second hardest jet (b)  in the signal
  and the SM backgrounds. }
\label{Figs:jet1}
\end{center}
\end{figure}

After the basic cuts to simulate the detector acceptance, we further employ optimized kinematical cuts to reduce the backgrounds  based on the kinematical differences between the signal and backgrounds.
 The signal events consist of five jets in the final sate.
These jets from the heavy boson $W_{H}$ decay tend to have larger $p_{T}$ than the jets in the background events. We order the jets by their $p_T$ and present the normalized $p_T$ distributions for the leading jet and the second jet in the signal events and background events in Fig.~\ref{Figs:jet1}. The model parameters are set as $m_{W_{H}} = 1.25$ TeV and $(x,r,\alpha)=(0.20,1,0.13)$ in the signal process.

From Fig.~\ref{Figs:jet1}.(a), we can see that the leading jet ($j_1$) in the signal events has
much harder $p_{T}$ distributions than the jet in the SM background events.
This is because the hardest jet in the signal is mainly the
 $b$ jet or the daughter-jet of the highly boosted top from $W_{H}\rightarrow t\bar{b}$ decay. Its $p_{T}$ spectrum peaks
around half of heavy boson mass.
However, the top quarks in the SM backgrounds are mainly produced in
the threshold region. Thus the jets from top quark decay in the backgrounds tend to
be soft.

\begin{figure}[htbp]
\begin{center}
\includegraphics [scale=0.5] {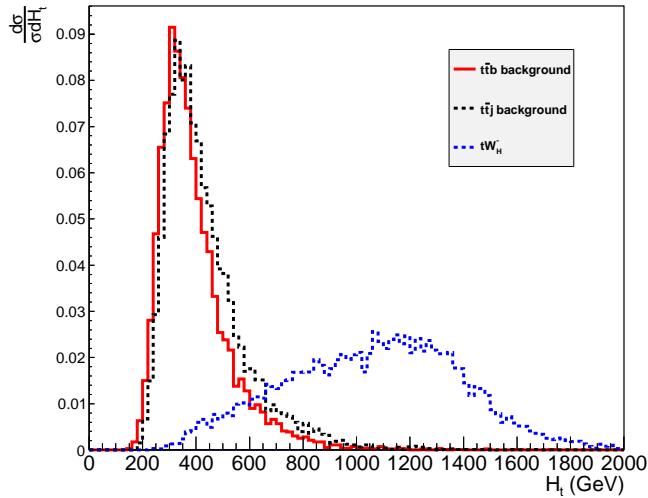}
\vspace{-0.2cm}\caption{Total transverse momentum $H_{t}$ distributions of the signal  and the SM backgrounds. }
\label{Figs:ht}
\end{center}
\end{figure}

Similar to the leading jet, the second jet ($j_{2}$) in the
signal is harder than that in the backgrounds as displayed in Fig.~\ref{Figs:jet1}.(b).
Furthermore, the normalized $H_{t}$ distributions, i.e.,
the scalar sum of the
$p_{T}$'s for all the visible particles in the final state,
 are shown in Fig.~\ref{Figs:ht}. The model parameters are set as $m_{W_{H}} = 1.25$ TeV and $(x,r,\alpha)=(0.20,1,0.13)$ in the signal process.

 To purify the
signal, a set of hard $p_{T}$  cuts are further adopted for $m_{W_{H}}= 1.25 $ TeV , based on above analysis, as follows:
\begin{subequations}
\begin{align}
   p_{T}(j_{1})\geq 250GeV,~~~~p_{T}(j_{2})\geq 140GeV,~~~~H_{t}=\sum^{5}_{1} p_{T}(j_{n})+ p_{T}(l^{+})~\geq 800GeV.
   \end{align}
   \label{eq15}
For $m_{W_{H}}=1.6$ TeV, a similar set of cuts
\begin{equation}
 p_{T}(j_{1})\geq 350GeV,~~~~p_{T}(j_{2})\geq 180GeV,~~H_{t}=\sum^{5}_{1} p_{T}(j_{n})+ p_{T}(l^{+})~\geq 1200GeV
   \end{equation}
   \end{subequations}
  are adopted.
These optimized cuts in Eq.~(\ref{eq15}) are referred to as cut II, and the efficiencies of these cuts are shown in Tables I and II.
\begin{figure}[htbp]
\begin{center}
\includegraphics [scale=0.50] {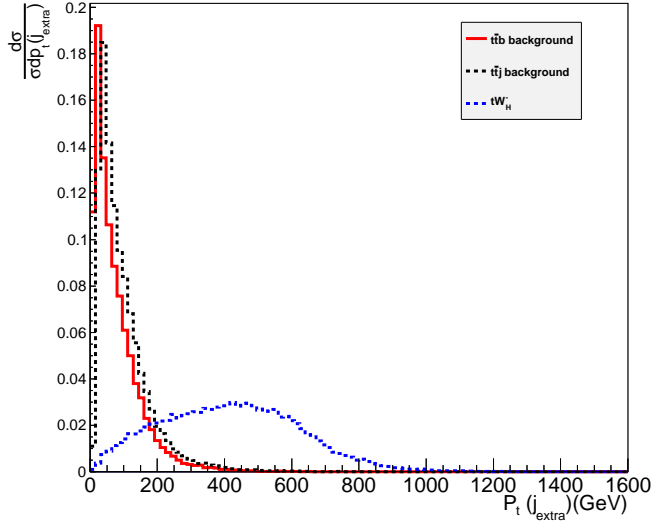}
\vspace{-0.2cm}\caption{Transverse momentum $p_{T}$ distributions of the extra jet in the signal and the SM backgrounds. }
\label{Eqs:extra}
\end{center}
\end{figure}

In order to suppress the $t\bar{t}j$
background, it is crucial to identify the extra jet (denoted as $j_{extra}$) produced in
association with the $t\bar{t}$ pair as a $b$ jet. Following the reconstruct scheme \cite{cqh}, we
apply the minimal $\chi^{2}$ -template method which based on the
$W$ boson and top quark masses to
pick out the extra jet,
\begin{eqnarray}
 \chi^{2}=\frac{(m_{t}-m_{jl\nu})^{2}}{\delta m^{2}_{t}}+
 \frac{(m_{t}-m_{jjj})^{2}}{\delta m^{2}_{t}}+
 \frac{(m_{W}-m_{jj})^{2}}{\delta m^{2}_{W}},
 \label{eq16}
\end{eqnarray}
where $\delta m_{t}$ and $\delta m_{W}$ are chosen to be 15 GeV and 10 GeV, respectively,
which account for the detector resolution capability. The $m_{W}$
and $m_{t}$ are taken as 80.4 GeV and 173.2 GeV, respectively.

In Fig.~\ref{Eqs:extra}, we present the normalized $p_{T}$ distributions of the extra jet for $W_{H}=1.25$ TeV and $(x,r,\alpha)=(0.20,1,0.13)$. The extra jet in association with $t\bar{t}$ in the SM backgrounds comes mainly from QCD radiation, while the extra jet in the signal is often  predominately from the heavy $W_{H}$ decay. Hence, it is obvious that  the extra jet in the signal has much harder $p_{T}$ distributions than the extra jet in the SM backgrounds.
Here, we further take cut on extra jet,
\begin{equation}
   p_{T}(j_{extra})\geq 200GeV.
   \label{eq17}
\end{equation}

Similarly, $p_{T}(j_{extra})\geq 400$ GeV  is adopted for  $m_{W_{H}}=1.6$ TeV.
After the extra jet is discriminated, we can further require it to be a $b$-jet.
Here, we choose the tagging efficiency of $b$-jet as $60\%$ and the mis-tagging
efficiency of a light quark jet and gluon jet as $1\%$ \cite{tag}.
This $b$ tagging cut could significantly suppress the $t\bar{t}j$ background.
\begin{figure}[htbp]
\begin{center}
\includegraphics [scale=0.5] {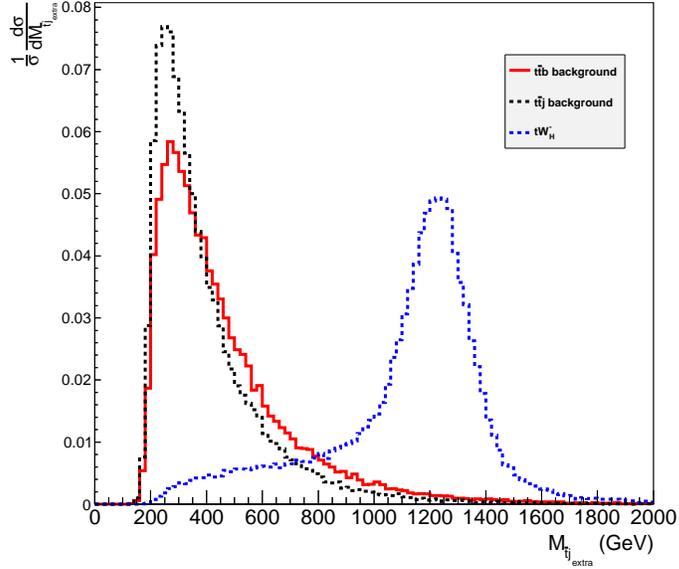}
\vspace{-0.2cm}\caption{ Invariant mass distribution of the reconstructed $\bar{t}$ and $j_{extra}$ in the signal and the SM backgrounds. }
\label{Eqs:mtj}
\end{center}
\end{figure}

\begin{table}
\begin{center}
\begin{tabular}{|c|c|c|c|c|c|}\hline
   &\multicolumn{3}{|c|}{singnal}
 &\multicolumn{2}{|c|}{backgrounds} \\ \hline
  &  $x=0.15$  &$ x=0.20$ &$x=0.25$ & $t\bar{t}b $  &$t\bar{t}j$ \\ \hline
 cuts I  &  $656 $  &$ 352 $ &$218$  &$2.27\times10^{4} $  &$2.78\times10^{6}$ \\ \hline
 cuts II   &  $453 $  &$ 247 $ &$152$  &$231$  &$2.86\times10^{4}$ \\ \hline
 b-tagging  &  $272 $  &$ 148 $&$91$  &$139$  &$286$ \\ \hline
 cuts III   &  $197 $  &$ 108 $&$68$  &$42$  &$18$ \\ \hline
 $S/\sqrt{B}$  &  $25.4 $  &$ 13.9 $ &$8.78$   &\multicolumn{2}{|c|}{---} \\ \hline
\end{tabular}
\caption{\small  The event numbers of the signal and backgrounds at the 14 TeV LHC
 with an integrated luminosity of 100 $fb^{-1}$  for $m_{W_{H}}= 1.25$ TeV and three values of $x$ after each performed cut. }
\end{center}
\end{table}
\begin{table}
\begin{center}
\begin{tabular}{|c|c|c|c|c|}\hline
Singnal & \multicolumn{4}{|c|}{$t\bar{t}b$} \\ \hline
M (TeV)      &\multicolumn{2}{|c|}{1.25}  &\multicolumn{2}{|c|}{1.60} \\ \hline
cuts I + II + III + b tagging   &\multicolumn{2}{|c|}{108} &\multicolumn{2}{|c|}{22} \\\hline\hline
 backgrounds  &  $t\bar{t}b$ & $t\bar{t}j$ &  $t\bar{t}b$ & $t\bar{t}j$  \\ \hline
cuts I + II + III + b tagging  &$42$  &$18$  &$2$  & $5$ \\ \hline
$S/\sqrt{B}$   & \multicolumn{2}{|c|}{13.9} & \multicolumn{2}{|c|}{8.32} \\ \hline
\end{tabular}
\caption{\small  The event numbers of the signal and backgrounds at the 14 TeV LHC
 with an integrated luminosity of 100 $fb^{-1}$  for $(x,r,\alpha)=(0.20,1,0.13)$ and three values of $m_{W_{H}}$ after each performed cut. }
\end{center}
\end{table}

After reconstructing the top pair and singling out the extra jet, it is easy to
reconstruct the $W_H$. We present the normalized invariant mass distribution of the
 $m_{\bar{t}j_{extra}}$ in  Fig.~\ref{Eqs:mtj}. It is shown that the signal distribution shows a sharp peak around the input value of $m_{W_{H}}$=1.25 TeV. However, the SM backgrounds exhibit a broad
spectrum and peak in the low mass region.
Thus we further require the invariant mass of the extra jet and $t$ quark or $\bar{t}$ quark to be around
the heavy boson $W_{H}$ mass window,
\begin{eqnarray}
  |m_{W_{H}}-m_{\bar{t}_{j_{extra}}}|< 200 GeV,\quad |m_{W_{H}}-m_{t_{j_{extra}}}|< 200 GeV.
  \label{eq18}
\end{eqnarray}
 The cuts in Eq.~(\ref{eq18}) can efficiently suppress the SM backgrounds while keep most of the signal.
These cuts in Eq.~(\ref{eq17}) and Eq.~(\ref{eq18}) are referred to as cut III, and the efficiency of these cuts are shown in Tables I and II.

As shown in Table I, the sets of cuts significantly suppress the backgrounds.
Supposing the integrated luminosity to be 100$fb^{-1}$ at
$\sqrt{s}=14 $ TeV, a large significance $S/\sqrt{B}=8.78$ (25.4) can be achieved for 1.25 TeV mass $W_{H}$ with $x=0.25$ (x=0.15).
In Table II, the case for $W_H$ with mass $m_{W_H}=1.6$ TeV and $(x,r,\alpha)=(0.20,1,0.13)$ are further considered. With a integrated luminosity of 100$fb^{-1}$,
a statistical significance $8.32 \sigma$ can be achieved for $W_H$ with mass $m_{W_H}=1.6$ TeV .


\section{Conclusions} \noindent

Many new physics scenarios beyond the SM predict the existence of new heavy gauge boson. The discovery of heavy charged gauge bosons will be the smoking gun of new gauge group and provide an important hint on electroweak symmetry breaking.

In this paper, focusing on the channel $pp\rightarrow tW_H \rightarrow t\bar{t}b$, we have studied
the potential for discovering the extra heavy gauge boson $W_{H}$ predicted in topflavor seesaw model at the LHC.
Studying the process $pp\rightarrow tW_H \rightarrow t\bar{t}b$ can independently test the $W_H\bar{t}b$ coupling and shed light on the flavor structure and the gauge structure of new physics models. In some new physics model \cite{9911266,cqh}, the couplings of SM fermions to new gauge boson $W_H$ are not universal. Especially, in the topflavor seesaw model, the couplings of $W_{H}$ with $\bar{t}b$ is enhanced by a factor of $1/x$ while the couplings of $W_H$ with light fermions are suppressed by a factor of $x$. After calculation, it is found that the cross section for $tW_H$ production can reach tens fb for $m_{W_H}$ in the mass range $1.2 \sim 1.7$ TeV. We further studied the different kinematic features of the signal $pp\rightarrow tW_H\rightarrow t\bar{t}b \rightarrow l\nu jjbbb$ and backgrounds.
Our study shows that it is possible to discover the signal of $W_H$ of topflavor seesaw model. The resonance peak in the invariant mass distribution of top quark and b-jet is a distinct signature of $W_H$ discovery. Taking the $m_{W_H}=1.6$ TeV and $(x,r,\alpha)=(0.20,1,0.13)$ as an example, the signal significance can reach $8.32 \sigma$
 at the LHC with $\sqrt{s} = 14$ TeV and luminosity $\LL= $ 100 $fb^{-1}$.
\\
\vspace{4mm} \textbf{Acknowledgments}\\
This work was supported in part by the National Natural Science Foundation of
China under Grants Nos.11275088,11175251, 11205023,  the Natural Science Foundation of the Liaoning Scientific Committee
(No. 201102114), Foundation of Liaoning Educational Committee (No. LT2011015) and the Natural Science Foundation of Dalian (No. 2013J21DW001).


\vspace{1.0cm}

\begin{thebibliography}\\



\bibitem{topcolorseesaw}
  B.~A.~Dobrescu and C.~T.~Hill,
  Phys.\ Rev.\ Lett.\  {\bf 81} (1998) 2634
  [hep-ph/9712319];
  R.~S.~Chivukula, B.~A.~Dobrescu, H.~Georgi and C.~T.~Hill,
  Phys.\ Rev.\ D {\bf 59}, 075003 (1999)
  [hep-ph/9809470];
  H.~-J.~He, C.~T.~Hill and T.~M.~P.~Tait,
  Phys.\ Rev.\ D {\bf 65}, 055006 (2002) [hep-ph/0108041].
\bibitem{9911266}
  H. J. He, T.M.P. Tait, C. P. Yuan. Phys. Rev. D 62,  011702 (2000) [hep-ph/9911266].


  \bibitem{ATLAS}
  G.~Aad {\it et al.}  [ATLAS Collaboration],
  Phys.\ Lett.\ B {\bf 716} (2012) 1
  [arXiv:1207.7214 [hep-ex]].

\bibitem{CMS}
  S.~Chatrchyan {\it et al.}  [CMS Collaboration],
  Phys.\ Lett.\ B {\bf 716} (2012) 30
  [arXiv:1207.7235 [hep-ex]].

  \bibitem{1304.2257}
X.~-F.~Wang, C.~Du and H.~-J.~He,
  Phys.\ Lett.\ B {\bf 723} (2013) 314
  [arXiv:1304.2257 [hep-ph]].
\bibitem{exd}
  N.~Arkani-Hamed, S.~Dimopoulos and G.~R.~Dvali,
  Phys.\ Lett.\ B {\bf 429} (1998) 263
  [hep-ph/9803315];
  L.~Randall and R.~Sundrum,
  Phys.\ Rev.\ Lett.\  {\bf 83}, 3370 (1999)[hep-ph/9905221];
K.~Agashe, A.~Delgado, M.~J.~May and R.~Sundrum,
  JHEP {\bf 0308} (2003) 050
  [hep-ph/0308036];
   R.~S.~Chivukula, D.~A.~Dicus, H.~-J.~He and S.~Nandi,
  Phys.\ Lett.\ B {\bf 562} (2003) 109
  [hep-ph/0302263].

\bibitem{lh}
   D. E. Kaplan, M. Schmaltz, JHEP 0310, 039 (2003) [hep-ph/0302049];
   N. Arkani-Hamed, A. G. Cohen, E. Katz,  A. E. Nelson. JHEP 0207, 034 (2002) [hep-ph/0206021];
   T. Han, H. E. Logan, B. McElrath,  L. T. Wang. Phys. Rev. D 67, 095004 (2003) [hep-ph/0301040].

 \bibitem{cqh}
    E.~L.~Berger, Q.~-H.~Cao, J.~-H.~Yu and C.~-P.~Yuan,
  Phys.\ Rev.\ D {\bf 84}, 095026 (2011)
  [arXiv:1108.3613 [hep-ph]].

 \bibitem{1111.5021}
    K.~S.~Babu, J.~Julio and Y.~Zhang,
  Nucl.\ Phys.\ B {\bf 858}, 468 (2012)
  [arXiv:1111.5021 [hep-ph]].

 \bibitem{g1312}
    T.~Jezo, M.~Klasen and I.~Schienbein,
  Phys.\ Rev.\ D {\bf 86}, 035005 (2012)
  [arXiv:1203.5314 [hep-ph]];
   C.~Du, H.~-J.~He, Y.~-P.~Kuang, B.~Zhang, N.~D.~Christensen, R.~S.~Chivukula and E.~H.~Simmons,
  Phys.\ Rev.\ D {\bf 86}, 095011 (2012)
  [arXiv:1206.6022 [hep-ph]];
  Q.~-H.~Cao, Z.~Li, J.~-H.~Yu and C.~P.~Yuan,
  Phys.\ Rev.\ D {\bf 86}, 095010 (2012)
  [arXiv:1205.3769 [hep-ph]];
     K.~Hsieh, K.~Schmitz, J.~-H.~Yu and C.~-P.~Yuan,
  Phys.\ Rev.\ D {\bf 82}, 035011 (2010)
  [arXiv:1003.3482 [hep-ph]].

  \bibitem{1011.5918}
     M.~Schmaltz and C.~Spethmann. JHEP {\bf 1107}, 046 (2011)  [arXiv:1011.5918 [hep-ph]].
  \bibitem{1111.1551}
  F.~Bach and T.~Ohl,
  Phys.\ Rev.\ D {\bf 85}, 015002 (2012)
  [arXiv:1111.1551 [hep-ph]].


\bibitem{CMSW'}
 S. Chatrchyan et al. [CMS Collaboration],
  Phys. Lett. B 718,  1229 (2013),[arXiv:1208.0956 [hep-ex]].
  \bibitem{ATLASW'}
 G. Aad et al. [ATLAS Collaboration], Eur. Phys. J. C 72,  2241 (2012), [arXiv:1209.4446 [hep-ex]].
  \bibitem{ATLASW'lv}
 G.~Aad {\it et al.}  [ATLAS Collaboration],
  Phys.\ Lett.\ B {\bf 705}, 28 (2011) [arXiv:1108.1316 [hep-ex]].
\bibitem{CMSW'lv}
  S.~Chatrchyan {\it et al.} [CMS Collaboration],
   JHEP {\bf 1208}, 023 (2012) [arXiv:1204.4764 [hep-ex]].
\bibitem{CMSW'WZ}
  S.~Chatrchyan {\it et al.} [CMS Collaboration],
  Phys.\ Rev.\ Lett.\  {\bf 109}, 141801 (2012) [arXiv:1206.0433 [hep-ex]].
\bibitem{ATLASW'tb}
  G.~Aad {\it et al.}  [ATLAS Collaboration],
  Phys.\ Rev.\ Lett.\  {\bf 109}, 081801 (2012) [arXiv:1205.1016 [hep-ex]].


 \bibitem{cteq}
  J. Pumplin, D. R. Stump, J. Huston, H. L. Lai, P. M. Nadolsky and W. K. Tung. JHEP 0207, 012 (2002) [arXiv:0201195 [hep-ph]].

  \bibitem{sm}
J. Beringer et al., [Particle Data Group], Phys. Rev. D 86, 010001 (2012).

  \bibitem{ylh}
  C. X. Yue, S. Yang,  L. H.~Wang. Europhys.\ Lett.\  {\bf 76}, 381 (2006) [hep-ph/0609107].

\bibitem{lrth}
   Y. B. Liu {\it et al.},   Commun. Theor. Phys. 49, 977  (2008).

\bibitem{shuo}
S. Yang and Q. S. Yan, JHEP 1202  074 (2012)  [arXiv:1111.4530 [hep-ph]].

\bibitem{mad}
 J. Alwall, M. Herquet, F. Maltoni, O. Mattelaer, T. Stelzer. JHEP 1106, 128 (2011)
[arXiv:1106.0522 [hep-ph]].

\bibitem{smear}
  G.~L.~Bayatian {\it et al.}  [CMS Collaboration],
  J.\ Phys.\ G {\bf 34}, 995 (2007).

 \bibitem{tag}
 G. Aad et al. [ATLAS Collaboration], (2009)  [arXiv:0901.0512 [hep-ex]].

\end{thebibliography}
\end{document}